\documentclass[journal]{IEEEtran}

\usepackage{graphicx}
\usepackage{amsmath,bbm,epsfig,amssymb,amsfonts,  amstext, verbatim,amsopn,cite,subfigure,multirow,multicol,lipsum}
\usepackage{balance}
\usepackage{url}
\usepackage{amsfonts}
\usepackage{epsfig}
\usepackage{setspace}
\usepackage{stmaryrd}
\usepackage{psfrag}	
\usepackage{multirow}
\usepackage{float}
\usepackage[process=auto]{pstool}
\usepackage{etoolbox}
\usepackage{algorithm}
\usepackage{algorithmic}
\usepackage{hyperref}

\usepackage{pifont}
\allowdisplaybreaks

\newtheorem{theo}{Theorem}

\newtheorem{remk}{Remark}

\newtheorem{corol}{Corollary}

\newtoggle{ConfVersion}

\toggletrue{ConfVersion}

\iftoggle{ConfVersion}{%
  
}{%
  
}

\EndPreamble
\begin{document}

\title{On the Design of Matched Filters for \\ Molecule Counting Receivers \vspace{-0.1cm}}

\author{Vahid Jamali, Arman Ahmadzadeh, and Robert Schober \vspace{-0.8cm}
}

\maketitle

\begin{abstract}
In this paper, we design matched filters for diffusive molecular communication systems taking into account the following impairments: \textit{signal-dependent} diffusion noise, \textit{inter-symbol interference} (ISI), and \textit{external interfering} molecules. The receiver counts the number of observed molecules several times within one symbol interval and employs linear filtering to detect the transmitted data. We derive the optimal matched filter by maximizing the expected signal-to-interference-plus-noise ratio of the decision variable. Moreover, we show that for the special case of an ISI-free channel, the matched filter reduces to a simple sum detector and a correlator for the channel impulse response for the diffusion noise-limited and (external) interference-limited regimes, respectively. Our simulation results reveal that the proposed matched filter considerably outperforms the benchmark schemes available in literature, especially when ISI is severe.
\end{abstract}


\section{Introduction}

Diffusive molecular communication (MC)  is a common  strategy for communication between nano-/microscale entities in nature such as bacteria, cells, and organelles (i.e., components of cells) \cite{CellBio}. Motivated by this observation, diffusive MC has been recently considered as a bio-inspired approach for communication between small-scale nodes where conventional wireless communications may be inefficient or even infeasible \cite{Nariman_Survey}.
However, establishing reliable diffusive MC is challenging due to many factors including the following impairments of the MC channel. First, diffusion is a random process and causes \textit{signal-dependent} noise.  Therefore, by releasing more molecules,  the variance of the diffusion noise increases as well. Second, since the MC channel is dispersive, the MC channel impulse response (CIR) may span several symbol intervals. This induces \textit{inter-symbol interference} (ISI) which impairs communication. Third, the receiver may be impaired by \textit{external interfering} molecules including multiuser interference (caused by other MC links) and environmental interference (originating from natural~sources)~\cite{TCOM_MC_CSI}.

Sequence detection was studied in \cite{Equ_MC,Akyl_Receiver_MC,Adam_OptReciever} to mitigate ISI in MC. However, sequence detection can be computationally complex specially for simple nano-machines which have limited computational capabilities.
Hence, symbol-by-symbol detection was advocated in \cite{Adam_OptReciever} where the receiver counts the number
of molecules several times within \textit{one symbol interval} and employs linear filtering to detect the transmitted data.
In particular, in \cite{Adam_OptReciever}, detection based on linear filtering was referred to as ``weighted sum detection"  and two \textit{heuristic options} were proposed for the weights, namely equal weights, i.e., a sum detector, and weights matched to the CIR, i.e., a~CIR~correlator. 
 
In this paper, we focus on the \textit{optimal design}  of the linear filter for symbol-by-symbol detection.  In particular, we take the three aforementioned  impairments into account and  derive the matched filter which maximizes the expected signal-to-interference-plus-noise ratio (SINR)\footnote{In \cite{Adam_OptReciever}, the CIR correlator was also referred to as ``matched filter"; however, the filter was not obtained based on a specific optimality criterion.}. To obtain further insight, we simplify the optimal matched filter for the case of an ISI-free channel where we show that the matched filter reduces to a simple sum detector and a CIR correlator for the diffusion noise-limited and (external) interference-limited regimes, respectively. Furthermore, we derive an approximate analytical expression for the bit error rate (BER). Finally, we provide simulation results to verify the analytical derivations and to evaluate the performance of the proposed~matched filter. 

\textit{Notations:} We use the following notations throughout this paper: $\mathsf{E}\{x\}$ and $\mathsf{Var}\{x\}$ denote the expectation and the variance of random variable (RV) $x$, respectively. Bold lower and upper case letters  denote vectors and matrices, respectively. Moreover, $\mathbf{A}^{\mathsf{T}}$ represents the transpose of matrix $\mathbf{A}$, $\mathsf{diag}(\mathbf{a})$ denotes a diagonal matrix with the elements of vector $\mathbf{a}$ as its main diagonal entries, and $\mathbf{1}_n$ is a vector of length $n$ whose elements~are~all~one.  In addition,  $\mathcal{P}(\lambda)$  denotes a Poisson~RV~with~mean~$\lambda$ and $\mathcal{N}(\mu,\sigma^2)$ represents a Gaussian random variable with mean $\mu$ and variance $\sigma^2$. Furthermore, $\kappa_{\max}(\cdot)$ and $\boldsymbol{\upsilon}_{\max}(\cdot)$ denote the maximum eigenvalue of a matrix and the corresponding eigenvector, respectively. Moreover, $\mathbb{Z}$ represents  the set of integer numbers, $n!$ is the factorial~of~$n$, and $\mathsf{Q}(\cdot)$ denotes the Q-function. 

\section{System Model}

\begin{figure}
  \centering
 \scalebox{0.4}{
\pstool[width=2\linewidth]{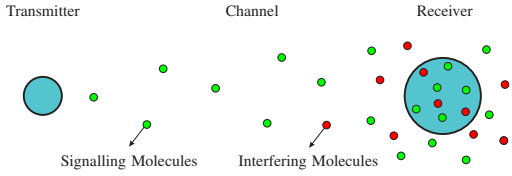}{
\psfrag{T}[c][c][1.5]{Transmitter}
\psfrag{R}[c][c][1.5]{Receiver}
\psfrag{C}[c][c][1.5]{Channel}
\psfrag{S}[c][c][1.5]{Signalling Molecules}
\psfrag{N}[c][c][1.5]{Interfering Molecules}
} } \vspace{-0.3cm}
\caption{Schematic illustration of the considered MC system.\vspace{-0.3cm}}
\label{Fig:TxRx}
\end{figure}

We consider an MC system  consisting of a transmitter, a channel, and a receiver, see Fig.~\ref{Fig:TxRx}. We employ on-off keying (OOK) modulation  where the transmitter releases zero or $N^{\mathrm{tx}}$ molecules at the beginning of the $k$-th symbol interval to send symbol $s[k]=0$ and $s[k]=1$, respectively \cite{Nariman_Survey}. The released molecules diffuse through the fluid medium between the transmitter and the receiver. Besides diffusion, signalling molecules may also be affected by other phenomena in the MC channel such as flow and molecule degradation. We assume synchronous transmission \cite{ICC2017_MC_Arxiv} where the receiver counts the number of observed molecules at sampling times $t_m=m\Delta t,\,\,m=1,\dots,M$, in a given symbol interval. Here, $\Delta t$ is the sampling interval and $M$ denotes the number of samples in each symbol interval. We further assume  that the MC channel has a memory of $L$ symbol intervals, i.e., the ISI in symbol interval $k$ originates from the $L-1$ previous  symbols.  The  number of  molecules counted at the receiver for sample $m$ in symbol interval $k$ is denoted by $r[k,m]$ and can be accurately modelled as a Poisson RV, see \cite{Yilmaz_Poiss,HamidJSAC,Adam_OptReciever,TCOM_MC_CSI}, i.e., 
\begin{align} \label{Eq:ChannelInOut}
  r[k,m]  \sim \mathcal{P}\left(\sum_{l=1}^L \left(\bar{c}_{\mathrm{s}}^{(l)}[m]s[k-l+1]\right)
  +\bar{c}_{\mathrm{i}}^{\mathrm{ext}}\right), 
\end{align}
where $\bar{c}^{(l)}_{\mathrm{s}}[m]$ is the  number of  molecules \textit{expected} to be observed at the receiver  at the $m$-th sampling time due to the release of $N^{\mathrm{tx}}$ molecules by the transmitter at the beginning of symbol interval $k-l+1$ and  $\bar{c}_{\mathrm{i}}^{\mathrm{ext}}$ is the \textit{expected} number of interfering noise molecules observed at the receiver which is assumed to be constant for all sampling times~\cite{TCOM_MC_CSI}. For \textit{given} transmit symbols $s[k'],\,\,\forall k'=k-L+1,\dots,k$, we assume that the observations $r[k,m]$ at different sampling times $m$~are~independent. 

\section{Optimal Filter Design}

In this section, we first derive an expression for the SINR at the output of a linear filter. Subsequently, we obtain the optimal matched filter which maximizes the SINR.

\subsection{SINR for Linear Filtering}

For deriving the SINR, we have to first identify the signal, noise, and interference terms in (\ref{Eq:ChannelInOut}). To this end, we rewrite~(\ref{Eq:ChannelInOut})~as
\begin{align} \label{Eq:SigNoiseInt}
  r[k,m]  = s[k]\bar{c}^{(1)}_{\mathrm{s}}[m]+ n[k,m] + I[k,m],
\end{align}
where $s[k]\bar{c}^{(1)}_{\mathrm{s}}[m]$ represents the desired signal, and $n[k,m]$ and $I[k,m]$ denote the diffusion noise and interference, respectively. Hereby, the diffusion noise follows distribution $n[k,m]\sim\bar{\mathcal{P}}(s[k]\bar{c}^{(1)}_{\mathrm{s}}[m])$  where the PDF of RV $X\sim\bar{\mathcal{P}}(\lambda)$ is given by
\begin{align} \label{Eq:NoisePDF}
f_{X}(x) = \frac{\lambda^{x+\lambda}\mathsf{exp}(-\lambda)}{(x+\lambda)!},\quad x\in\mathcal{X},
\end{align}
and $\mathcal{X}=\{x\in\mathbb{Z}| x\geq -\lambda\}$. For RV $X$, we have $\mathsf{E}\{X\}=0$ and $\mathsf{Var}\{X\}=\lambda$.  In other words, $n[k,m]$ is equivalent to a Poisson RV whose mean has been subtracted. We note that as $\lambda\to\infty$, we obtain $\bar{\mathcal{P}}(\lambda)\to\mathcal{N}(0,\lambda)$. Moreover, $I[k,m]$ is a Poisson RV, i.e., $I[k,m]\sim\mathcal{P}\left(\bar{c}_{\mathrm{i}}[k,m]\right)$ where $\bar{c}_{\mathrm{i}}[k,m] = \bar{c}_{\mathrm{i}}^{\mathrm{isi}}[k,m]+\bar{c}_{\mathrm{i}}^{\mathrm{ext}}$ and $\bar{c}_{\mathrm{i}}^{\mathrm{isi}}[k,m]=\sum_{l=2}^L \bar{c}_{\mathrm{s}}^{(l)}[m]s[k-l+1]$.

The structure of the received signal $r[k,m]$ in (\ref{Eq:SigNoiseInt}) is similar to that in an additive white Gaussian noise (AWGN) channel with interference for conventional wireless communication systems. The main differences are that, here, diffusion noise and interference are not Gaussian distributed and the diffusion noise is signal dependent since  its variance depends on the desired signal $s[k]$, i.e., $\mathsf{Var}\{n[k,m]\}=\bar{c}^{(1)}_{\mathrm{s}}[m]s[k]$.

For complexity reasons, the decision variable is computed based on the observations in one symbol interval only. Hence, assuming linear processing, the decision variable at the output of the linear filter, $y[k]$, is obtained as
\begin{align} \label{Eq:LinearFilter}
y[k] = \mathbf{f}^\mathsf{T}\mathbf{r}[k] 
=  \mathbf{f}^\mathsf{T}\left(s[k]\bar{\mathbf{c}}_{\mathrm{s}}+\mathbf{n}[k]+\mathbf{I}[k]\right),
\end{align}
where $\mathbf{f}=[f[1],f[2],\dots,f[M]]^\mathsf{T}$ is the linear filter in the $k$-th symbol interval and $\mathbf{r}[k]$, $\bar{\mathbf{c}}_{\mathrm{s}}$, $\mathbf{n}[k]$, and $\mathbf{I}[k]$ are vectors containing the $r[k,m]$, $\bar{c}^{(1)}_{\mathrm{s}}[m]$, $n[k,m]$, and $I[k,m]$  for all sampling times $m$ in symbol interval $k$, respectively.
The SINR at the output of the filter  is defined as  
\begin{align} \label{Eq:SINR}
\mathsf{SINR} &=\frac{\mathsf{E}\left\{\big(\mathbf{f}^\mathsf{T}s[k]\bar{\mathbf{c}}_{\mathrm{s}}\big)^2\right\}}
{\mathsf{Var}\left\{\mathbf{f}^\mathsf{T}\mathbf{n}[k]\right\}
+\mathsf{Var}\left\{\mathbf{f}^\mathsf{T}\mathbf{I}[k]\right\}} \nonumber \\
&\overset{(a)}{=}  \frac{0.5\mathbf{f}^\mathsf{T}\bar{\mathbf{c}}_{\mathrm{s}}\bar{\mathbf{c}}_{\mathrm{s}}^\mathsf{T}\mathbf{f}}
{\mathbf{f}^\mathsf{T} \left(0.5\mathsf{diag}\left\{\bar{\mathbf{c}}_{\mathrm{s}}\right\} +\bar{\mathbf{C}}_{\mathrm{i}}\right) \mathbf{f}},
\end{align}
where $\bar{\mathbf{C}}_{\mathrm{i}}$  is the covariance matrix of $\mathbf{I}[k]$.  For equality $(a)$, we assumed equiprobable OOK symbols, and exploited the mutual independence of the diffusion noise for different sampling times, and used (\ref{Eq:NoisePDF}) for obtaining the variance of the diffusion noise. Let vector $\mathbf{s}\in\{0,1\}^{L-1}$ be a possible realization of $[s[k-L+1],\dots,s[k-1]]^{\mathsf{T}}$, and let $\bar{c}_{\mathrm{i}}[m|\mathbf{s}]$ denote the mean of the interference term, $\bar{c}_{\mathrm{i}}[k,m]$, conditioned on a given $\mathbf{s}$. Exploiting the properties of the Poisson distribution, the element in the $m$-th row and the $m'$-th column of $\bar{\mathbf{C}}_{\mathrm{i}}$ is obtained as
\begin{IEEEeqnarray}{lll}  \label{Eq:CovMat}
\bar{\mathbf{C}}_{\mathrm{i}}(m,m') = \frac{1}{2^{L-1}} \times \\
\begin{cases}
 \sum_{\forall \mathbf{s}} \bar{c}_{\mathrm{i}}[m|\mathbf{s}]\bar{c}_{\mathrm{i}}[m'|\mathbf{s}]  \\
 - \frac{1}{2^{L-1}}
\sum_{\forall \mathbf{s}} \bar{c}_{\mathrm{i}}[m|\mathbf{s}] 
\sum_{\forall \mathbf{s}} \bar{c}_{\mathrm{i}}[m'|\mathbf{s}],\qquad\qquad\qquad  \mathrm{if}\,\, m\neq m' \\
\sum_{\forall \mathbf{s}}
\bar{c}^2_{\mathrm{i}}[m|\mathbf{s}]+\bar{c}_{\mathrm{i}}[m|\mathbf{s}]  - \frac{1}{2^{L-1}}
\left(\sum_{\forall \mathbf{s}} \bar{c}_{\mathrm{i}}[m|\mathbf{s}] \right)^2,\,\,\mathrm{otherwise.} \hspace{-0.5cm}
\end{cases} \nonumber
\end{IEEEeqnarray}
Note that for an ISI-free MC channel,  $\bar{\mathbf{C}}_{\mathrm{i}}$ is a diagonal matrix.

\subsection{Matched Filter}

Our goal is to obtain the optimal matched filter $\mathbf{f}$ which maximizes the SINR in (\ref{Eq:SINR}). In particular, we have the following optimization problem
\begin{align} \label{Eq:OptProb}
\mathbf{f}^{\mathrm{opt}}=\underset{\mathbf{f}}{\mathrm{argmax}} \,\, \frac{\mathbf{f}^\mathsf{T}\bar{\mathbf{c}}_{\mathrm{s}}\bar{\mathbf{c}}_{\mathrm{s}}^\mathsf{T}\mathbf{f}}
{\mathbf{f}^\mathsf{T}\left(0.5\mathsf{diag}\left\{\bar{\mathbf{c}}_{\mathrm{s}}\right\}+\bar{\mathbf{C}}_{\mathrm{i}}\right)\mathbf{f}}.
\end{align}
A closed-form solution of the above optimization problem is given in the following theorem.
\begin{theo}\label{Theo:MF}
The optimal matched filter as the solution of (\ref{Eq:OptProb}) and the resulting maximum SINR are given by
\begin{IEEEeqnarray}{rl} 
\mathbf{f}^{\mathrm{opt}} &= \left( 0.5\mathsf{diag}\left\{\bar{\mathbf{c}}_{\mathrm{s}}\right\}+\bar{\mathbf{C}}_{\mathrm{i}} \right)^{-1} \bar{\mathbf{c}}_{\mathrm{s}} \nonumber \\
&\overset{(a)}{=} \left[\frac{\bar{c}_{\mathrm{s}}[1]}{0.5\bar{c}_{\mathrm{s}}[1]+\bar{c}_{\mathrm{i}}^{\mathrm{ext}}},\dots,\frac{\bar{c}_{\mathrm{s}}[M]}{0.5\bar{c}_{\mathrm{s}}[M]+\bar{c}_{\mathrm{i}}^{\mathrm{ext}}}\right]^{\mathsf{T}} \,\, \text{and}\quad \label{Eq:OptMF} \\
\mathsf{SINR}^{\mathrm{opt}} &= 0.5\bar{\mathbf{c}}_{\mathrm{s}}^{\mathsf{T}} \left( 0.5\mathsf{diag}\left\{\bar{\mathbf{c}}_{\mathrm{s}}\right\}+\bar{\mathbf{C}}_{\mathrm{i}} \right)^{-1} \bar{\mathbf{c}}_{\mathrm{s}} \nonumber \\
& \overset{(b)}{=}  \sum_{m=1}^M \frac{0.5\bar{c}^2_{\mathrm{s}}[m]}{0.5\bar{c}_{\mathrm{s}}[m]+\bar{c}_{\mathrm{i}}^{\mathrm{ext}}}
\triangleq \sum_{m=1}^M \mathsf{SINR}[k,m] ,\quad
\end{IEEEeqnarray}
respectively, where $\mathsf{SINR}[k,m]=\frac{0.5\bar{c}^2_{\mathrm{s}}[m]}{0.5\bar{c}_{\mathrm{s}}[m]+\bar{c}_{\mathrm{i}}^{\mathrm{ext}}}$ and equalities $(a)$ and $(b)$ hold for ISI-free channels.
\end{theo}
\begin{IEEEproof}
The proof is provided in the Appendix.
\end{IEEEproof}

\begin{remk}
Theorem~\ref{Theo:MF} reveals that the optimal matched filter can be given in closed form as in  (\ref{Eq:OptMF}). Furthermore, the proposed matched filter  requires only linear operations with respect to the observation samples, i.e., $y[k] = \mathbf{f}^\mathsf{T}\mathbf{r}[k]$. Interestingly, in nature,  a single neuron is able to perform summation and multiplication operations which suggests that the operations required for calculation of $y[k]$ can be implemented with (synthetic) biological systems  \cite{NeoronSumProd}.
\end{remk}

\begin{remk} We note that the optimal matched filter depends only on the channel state information (CSI) of the MC channel which comprises $\bar{c}_{\mathrm{s}}^{(l)}[m],\,\,\forall l,m,$ and $\bar{c}_{\mathrm{i}}^{\mathrm{ext}}$. This CSI can be obtained offline \cite{TCOM_MC_CSI}, and then be used for online detection. Moreover, it has to be updated only when the MC channel changes. For the case when CSI is not available, non-coherent detection schemes can be considered, see e.g. \cite{NonCoherent_MC,NanoCOM16}.
\end{remk}


\subsection{Special Cases}
In this subsection, we focus on an ISI-free MC channel. Therefore, the mean of the interference does not depend on the previous symbols and hence it is constant for all sample times, i.e., $\bar{c}_{\mathrm{i}}[k,m]=\bar{c}_{\mathrm{i}}^{\mathrm{ext}},\,\,\forall k,m$. Thereby, we consider two special cases, namely \textit{i)} the noise-limited regime where the diffusion noise is dominant over the interference, i.e., when all elements of $\bar{\mathbf{c}}_{\mathrm{s}}$ are much larger than $\bar{c}_{\mathrm{i}}^{\mathrm{ext}}$, and \textit{ii)} the interference-limited regime where the interference is dominant over the diffusion noise, i.e., when all elements of $\bar{\mathbf{c}}_{\mathrm{s}}$ are much smaller than $\bar{c}_{\mathrm{i}}^{\mathrm{ext}}$. 
Considering that diffusion noise is signal dependent, the noise- and interference-limited regimes are equivalent to very low and very high SINRs, respectively. The following corollary provides simplified matched filters for these two cases.

\begin{corol}\label{Corol:MF_Special}
For the noise- and interference-limited regimes, the optimal matched filter in (\ref{Eq:OptMF}) simplifies to 
\begin{IEEEeqnarray}{ll}\label{Eq:MF_Special} 
\mathbf{f}^{\mathrm{opt}}= 
\begin{cases}
\mathbf{1}_M,\,\,&\text{Noise-limited regime}  \\
\bar{\mathbf{c}}_{\mathrm{s}},\,\,&\text{Interference-limited regime.}
\end{cases}\quad
\end{IEEEeqnarray}
\end{corol} 
\begin{IEEEproof}
For the noise-limited regime, simplifying (\ref{Eq:OptMF}) using $\bar{c}_{\mathrm{s}}[m]\gg\bar{c}_{\mathrm{i}}^{\mathrm{ext}}$ leads to $\mathbf{f}^{\mathrm{opt}}=2\mathbf{1}_M$ and for the interference-limited regime, simplifying (\ref{Eq:OptMF}) using $\bar{c}_{\mathrm{s}}[m]\ll \bar{c}_{\mathrm{i}}^{\mathrm{ext}}$ leads to  $\mathbf{f}^{\mathrm{opt}}= \frac{1}{\bar{c}_{\mathrm{i}}^{\mathrm{ext}}}\bar{\mathbf{c}}_{\mathrm{s}}$. However, since the factors $2$ and $\frac{1}{\bar{c}_{\mathrm{i}}^{\mathrm{ext}}}$~are~constant, they do not affect the SINR in (\ref{Eq:SINR}) and hence,~$\mathbf{f}^{\mathrm{opt}}= \mathbf{1}_M$ and $\mathbf{f}^{\mathrm{opt}}= \bar{\mathbf{c}}_{\mathrm{s}}$ are also solutions of (\ref{Eq:OptProb}) for the noise- and interference limited regimes, respectively. This completes~the~proof.
\end{IEEEproof}

Corollary~\ref{Corol:MF_Special} reveals an interesting insight for the two considered extreme cases. In particular, for high SINRs, the optimal matched filter reduces to the simple sum detector which does not require  the CSI of the MC channel. On the other hand, for low SINRs, the optimal matched filter is the well-known CIR correlator.


\subsection{BER Analysis}

We adopt the following simple threshold detector 
\begin{IEEEeqnarray}{ll}\label{Eq:Detector} 
\hat{s}[k]=
\begin{cases}
1,\,\,&\mathrm{if}\,\, y[k]\geq \xi\\
0, &\mathrm{otherwise},
\end{cases} \quad
\end{IEEEeqnarray}
where $\xi$ is the detection threshold. Recall that $r[k,m]$ is a Poisson RV, cf.~(\ref{Eq:ChannelInOut}), and hence, $y[k] = \mathbf{f}^\mathsf{T}\mathbf{r}[k] $ is a weighted sum of Poisson RVs. Unfortunately, deriving the exact BER based on the distribution of $y[k]$ does not lead to an insightful expression. Therefore, given $\mathbf{s}[k]=\mathbf{s}$ and $s[k]=i$, we consider the following approximation
\begin{IEEEeqnarray}{ll}\label{Eq:Gaussian} 
y[k]\sim\mathcal{N}\left(\mu_i(\mathbf{s}),\sigma_i^2(\mathbf{s})\right), 
\end{IEEEeqnarray}
where $\mu_0(\mathbf{s})=\mathbf{f}^\mathsf{T}\bar{\mathbf{c}}_{\mathrm{i}}(\mathbf{s})$, 
$\mu_1(\mathbf{s})=\mathbf{f}^\mathsf{T}\left(\bar{\mathbf{c}}_{\mathrm{s}}+\bar{\mathbf{c}}_{\mathrm{i}}(\mathbf{s})\right)$,
$\sigma_0^2(\mathbf{s}) = \mathbf{f}^\mathsf{T}  \mathsf{diag}\{\bar{\mathbf{c}}_{\mathrm{i}}(\mathbf{s})\}  \mathbf{f}$, 
$\sigma_1^2(\mathbf{s}) = \mathbf{f}^\mathsf{T}  \mathsf{diag}\{\bar{\mathbf{c}}_{\mathrm{s}}+\bar{\mathbf{c}}_{\mathrm{i}}(\mathbf{s})\}  \mathbf{f}$,
and $\bar{\mathbf{c}}_{\mathrm{i}}(\mathbf{s})=\left[\bar{c}_{\mathrm{i}}[1|\mathbf{s}],\dots,\bar{c}_{\mathrm{i}}[M|\mathbf{s}]\right]^{\mathsf{T}}$. The above approximation becomes valid when $\mu_i(\mathbf{s})$ is large.

  Based on the approximation in (\ref{Eq:Gaussian}), the BER of the threshold detector in (\ref{Eq:Detector}) is obtained as
\begin{IEEEeqnarray}{ll}\label{Eq:BER} 
P_e[k] &= \sum_{\forall\mathbf{s}} P_e^{c}[k|\mathbf{s}]\mathsf{Pr}\left\{\mathbf{s}[k]=\mathbf{s}\right\}=\frac{1}{2^{L-1}}\sum_{\forall\mathbf{s}} P_e^{c}[k|\mathbf{s}],\quad\,\,
\end{IEEEeqnarray}
where $P_e^c[k|\mathbf{s}]=\mathsf{Pr}\left\{\hat{s}[k]\neq s[k]|\mathbf{s}[k]=\mathbf{s}\right\}$ is the error probability conditioned on the previous symbols  and is given by
\begin{IEEEeqnarray}{ll}\label{Eq:BER_condition} 
P_e^c[k|\mathbf{s}] &= \mathsf{Pr}\left\{y[k]< \xi|s[k]=1,\mathbf{s} \right\} \mathsf{Pr}\left\{s[k]=1\right\} \nonumber \\
&\,\,\,\,+ \mathsf{Pr}\left\{y[k]\geq \xi|s[k]=0,\mathbf{s} \right\}\mathsf{Pr}\left\{s[k]=0\right\} \nonumber \\
& = \left[1-\mathsf{Q}\left(\frac{\xi-\mu_1\left(\mathbf{s}\right)}{\sigma_1\left(\mathbf{s}\right)}\right)+\mathsf{Q}\left(\frac{\xi-\mu_0\left(\mathbf{s}\right)}{\sigma_0\left(\mathbf{s}\right)}\right)\right].\quad
\end{IEEEeqnarray}

\begin{remk}
Note that instead of optimizing $\mathbf{f}$ for SINR maximization as in (\ref{Eq:OptProb}), one may wish to optimize $\mathbf{f}$ to directly minimize the approximate BER  in (\ref{Eq:BER}). However, as can be seen from (\ref{Eq:BER}), the dependence of the BER on $\mathbf{f}$ through $\mu_0$, $\mu_1$, and the Q-function makes such a BER minimization problem in general intractable. 
\end{remk}

\begin{figure*}[!tbp]
  \centering
  \begin{minipage}[b]{0.3\textwidth}
  \hspace{-0.5cm}
  \centering
\resizebox{1.1\linewidth}{!}{\psfragfig{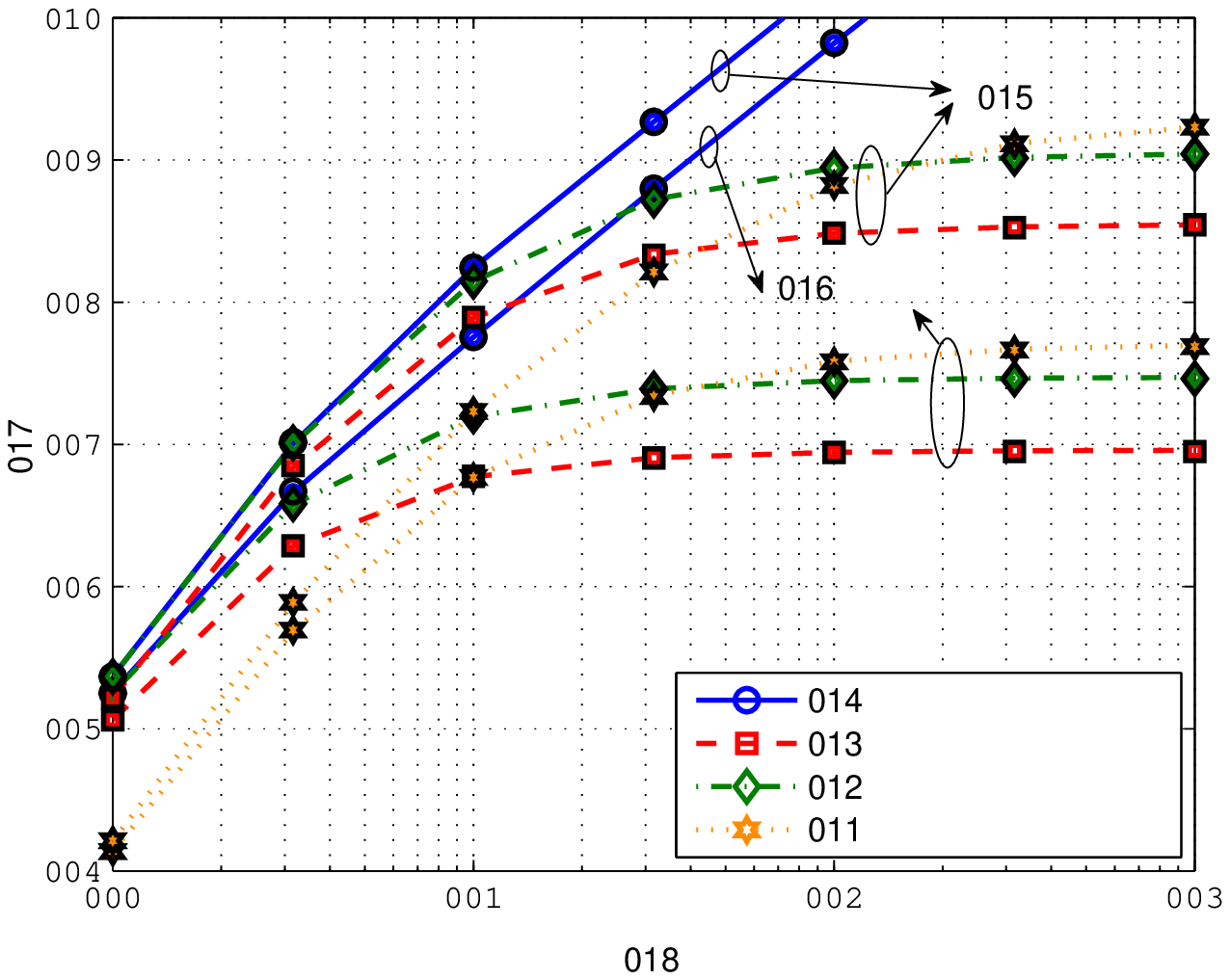}} \vspace{-0.7cm} 
\caption{SINR vs. $N^{\mathrm{tx}}$ for $\overline{T}^{\mathrm{symb}}=\{1.5,3\}$. \vspace{-0.2cm} }
\label{Fig:SINR_Ntx}
  \end{minipage}
    \hfill
  \begin{minipage}[b]{0.3\textwidth}
  \hspace{-0.5cm}
  \centering
\resizebox{1.1\linewidth}{!}{\psfragfig{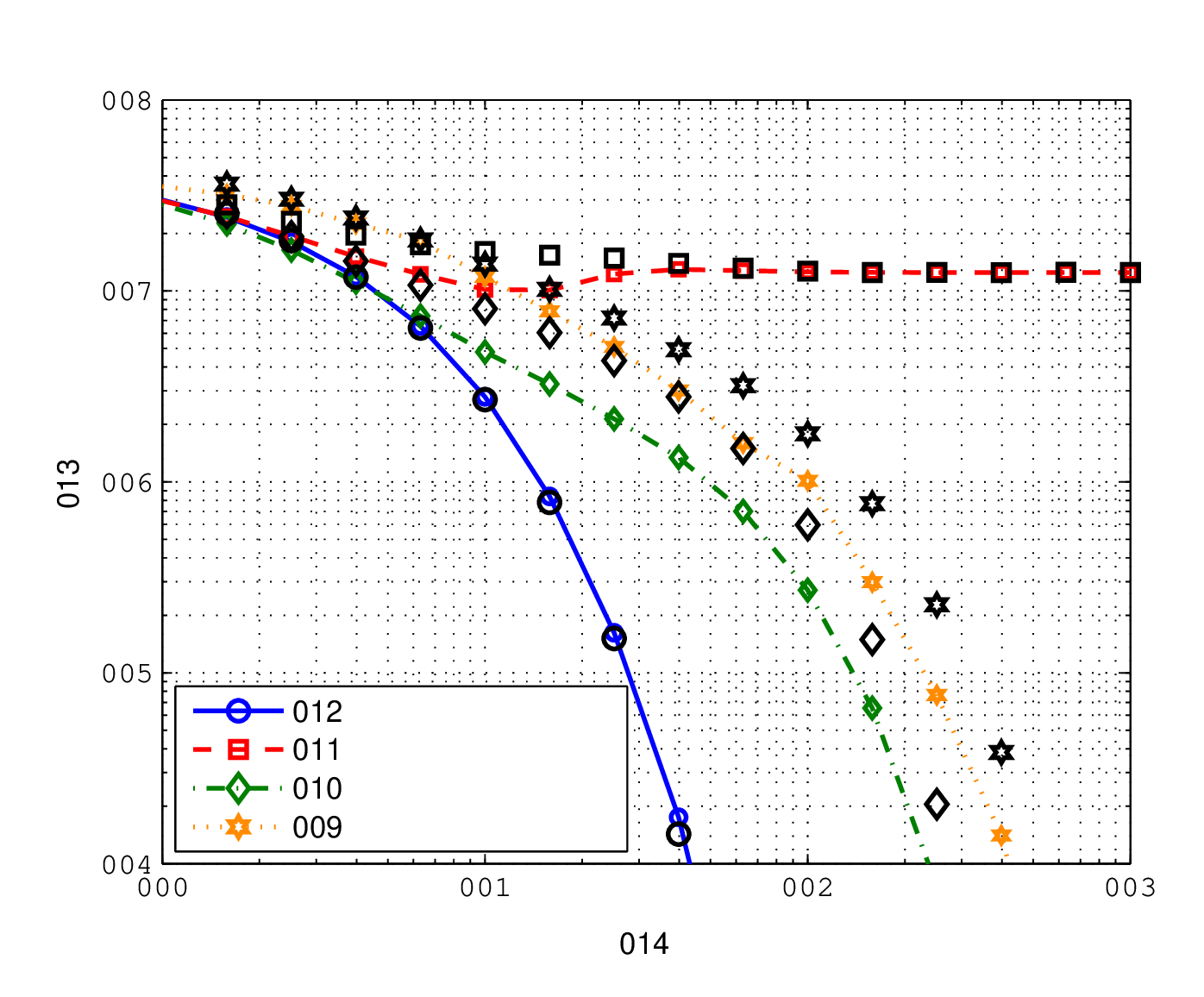}} \vspace{-0.7cm} 
\caption{BER vs. $N^{\mathrm{tx}}$  for $\overline{T}^{\mathrm{symb}}=1.5$. \vspace{-0.2cm} }
\label{Fig:BER_Ntx}
  \end{minipage}
      \hfill
  \begin{minipage}[b]{0.3\textwidth}
  \hspace{-0.5cm}
  \centering
\resizebox{1.1\linewidth}{!}{\psfragfig{}} \vspace{-0.7cm} 
\caption{$\mathbf{f}^{\mathrm{opt}}$ vs. $N^{\mathrm{tx}}$ for $\overline{T}^{\mathrm{symb}}=1.5$. \vspace{-0.2cm} }
\label{Fig:Filter_Ntx}
  \end{minipage}
  \hfill
  \begin{minipage}[b]{0.1\textwidth}
  \end{minipage} \vspace{-0.4cm}
\end{figure*}

\begin{table}
\label{Table:Parameter}
\caption{Default Values of the System Parameters \cite{TCOM_MC_CSI,Adam_OptReciever}. \vspace{-0.2cm}} 
\begin{center}
\scalebox{0.65}
{
\begin{tabular}{|| c | c | c ||}
  \hline 
  Variable & Definition & Value \\ \hline \hline
       $V^{\mathrm{rx}}$ & Receiver volume   & $\frac{4}{3}\pi 50^3$  ${\text{nm}}^3$ \\  \hline
               $d$ &  Distance between the transmitter and the receiver  & $500$ nm\\ \hline 
         $D$ &  Diffusion coefficient for the signaling molecule & $4.3\times 10^{-10}$ $\text{m}^2\cdot\text{s}^{-1}$\\ \hline          
       $\bar{c}_{\mathrm{e}}$ &  Enzyme concentration  & $10^{5}$ $\text{molecule}\cdot\mu\text{m}^3$ \\   \hline
       $\kappa$ &    Rate of molecule degradation reaction & $2\times10^{-19}$ $\text{s}^{-1}$ \\ \hline 
       $(v_{\parallel},v_{\perp})$ &  Components of flow velocity   & $(10^{-3},10^{-3})$ $\text{m}\cdot\text{s}^{-1}$ \\ \hline
\end{tabular}
}
\end{center}\vspace{-0.5cm}
\end{table}

\section{Performance Evaluation}

For performance evaluation, we consider a simple transparent receiver and a three-dimensional unbounded environment with uniform flow where enzyme molecules are uniformly present and degrade the signalling molecules \cite{Adam_OptReciever}\footnote{We emphasize that the application of the proposed matched filter detector is not limited to the example MC channel used here for simulation. In particular, the calculation of the proposed matched filter requires CSI which, for the considered example MC channel model, can be obtained in closed form using (\ref{Eq:Consentration}). For general MC systems, CSI can be obtained using e.g. training-based channel~estimators~\cite{TCOM_MC_CSI}.}. Thereby, the \textit{expected} number of molecules observed at the receiver as a function of time is given by
\begin{align} \label{Eq:Consentration}
\hspace{-0.3cm} \bar{c}_{\mathrm{s}}(t) \hspace{-0.05cm}=\hspace{-0.05cm} \frac{N^{\mathrm{tx}}V^{\mathrm{rx}}}{(4\pi D t)^{3/2}} \mathrm{exp}\left(\hspace{-0.1cm}-\kappa\bar{c}_{\mathrm{e}}t-\frac{(d-v_{\parallel}t)^2+(v_{\perp}t)^2}{4Dt}\right),\hspace{-0.15cm}
\end{align}
where the definition of the involved variables and their default values are provided in Table~I \cite{CellBio,Adam_OptReciever}. We employ $\bar{c}_{\mathrm{s}}(t)$ to obtain both $\bar{\mathbf{c}}_{\mathrm{s}}$ and the ISI component of $\bar{\mathbf{C}}_{\mathrm{i}}$ and adopt $\bar{c}_{\mathrm{i}}^{\mathrm{ext}}=2$. Let us define $T^{\mathrm{ref}}={\mathrm{argmax}}_t\,\,\bar{c}_{\mathrm{s}}(t)$ as a reference time, i.e., $T^{\mathrm{ref}}=0.176$ ms for the parameter values in Table~I. Then, we adopt a normalized sampling interval of $\overline{\Delta t}= \frac{\Delta t}{T^{\mathrm{ref}}}=0.25$ and normalized symbol durations of $\overline{T}^{\mathrm{symb}}= \frac{T^{\mathrm{symb}}}{ T^{\mathrm{ref}}}\in\{1.5,3\}$\footnote{We note that the number of samples per symbol cannot be arbitrarily increased, as for a fixed sampling time, the number of samples per symbol is limited by the symbol duration, and for a given symbol duration, the sampling interval should be large enough such that the independence of consecutive samples is ensured~\cite{Adam_OptReciever}.}. Moreover, we employ $M=6$ samples in each symbol interval for detection and assume an MC channel with $L=3$ taps. Finally, our simulation results are obtained by averaging over $10^{6}$ channel realizations, where each channel realization is generated as a Poisson RV with the mean as given in (\ref{Eq:ChannelInOut}). As benchmark schemes, we consider the sum filter, $\mathbf{f}=\mathbf{1}_M$, and the CIR correlator, $\mathbf{f}=\bar{\mathbf{c}}_{\mathrm{s}}$, from \cite{Adam_OptReciever}, and a peak detector, which employs only one sample at $T^{\mathrm{ref}}$ after the start of the symbol interval \cite{Nariman_Survey}.

In Fig.~\ref{Fig:SINR_Ntx}, we show the SINR vs. $N^{\mathrm{tx}}$ where the simulation results are shown by black markers and the analytical results from (\ref{Eq:SINR}) are shown as colored lines. We observe an excellent match between the analytical and simulation results. Moreover, we observe that as $N^{\mathrm{tx}}$ increases, the SINR achieved by all filters increases. Nevertheless, the SINRs of the benchmark schemes saturate at certain levels as $N^{\mathrm{tx}}\to\infty$ whereas the SINR of the matched filter does not saturate for the considered values of $N^{\mathrm{tx}}$. The saturation of the SINRs of the benchmark schemes is due to the strong ISI induced by increasing $N^{\mathrm{tx}}$. In fact, the aforementioned destructive effect of the ISI is more severe for $\overline{T}^{\mathrm{symb}}=1.5$ than for $\overline{T}^{\mathrm{symb}}=3$ as the symbol duration is smaller in the former case.

In Fig.~\ref{Fig:BER_Ntx}, we plot the BER vs. $N^{\mathrm{tx}}$ for $\overline{T}^{\mathrm{symb}}=1.5$. Again, the simulation results are shown by black markers and the analytical approximation results from (\ref{Eq:BER}) and (\ref{Eq:BER_condition}) are shown as colored lines. From Fig.~\ref{Fig:BER_Ntx}, we observe that although the proposed approximation follows the trend of the BER curves, it is not tight  for the correlator and peak filters. This mismatch is mainly due to the instances where $\mathbf{s}=[0,0]^{\mathsf{T}}$ holds. In particular, for these instances, we obtain the relatively small value $\mu_0(\mathbf{s})=\bar{c}_{\mathrm{i}}^{\mathrm{ext}}=2$ for which the corresponding Gaussian approximation in (\ref{Eq:Gaussian}) is not accurate. Fig.~\ref{Fig:BER_Ntx} reveals that as expected from the SINR analysis in Fig.~\ref{Fig:SINR_Ntx}, the proposed matched filter considerably outperforms the benchmark schemes in terms of BER.

To obtain further insight, in Fig.~\ref{Fig:Filter_Ntx}, we plot the values of the matched filter coefficients vs. $N^{\mathrm{tx}}$ for $\overline{T}^{\mathrm{symb}}=1.5$. Interestingly, the first two coefficients of the matched filter assume negative values for large $N^{\mathrm{tx}}$. The reason for this behavior is that the first samples in each symbol interval experience the most severe ISI and since the matched filter exploits the statistics of the ISI via $\bar{\mathbf{C}}_{\mathrm{i}}$,  it attempts to reduce the ISI by assigning negative values to $f^{\mathrm{opt}}[1]$ and $f^{\mathrm{opt}}[2]$.

%

\vspace{-0.2cm}

\appendix

We exploit the Rayleigh quotient inequality \cite{Matrix_Book,WPMC_2011}, i.e., 
\begin{align} 
\frac{\mathbf{x}^{\mathsf{T}}\mathbf{A}\mathbf{x}}{\mathbf{x}^{\mathsf{T}}\mathbf{B}\mathbf{x}} 
\leq \kappa_{\max}\left(\mathbf{B}^{-1}\mathbf{A}\right),
\end{align}
where $\mathbf{A}$ is a Hermitian matrix, $\mathbf{B}$ is a Hermitian positive-definite matrix, and the inequality holds with equality for $\mathbf{x}=\boldsymbol{\upsilon}_{\max}\left(\mathbf{B}^{-1}\mathbf{A}\right)$. Therefore, we obtain the optimal filter as
\begin{align} 
 \mathbf{f}^{\mathrm{opt}}=\boldsymbol{\upsilon}_{\max}\left( \left( 0.5\mathsf{diag}\left\{\bar{\mathbf{c}}_{\mathrm{s}}\right\}+\bar{\mathbf{C}}_{\mathrm{i}} \right)^{-1}
 \bar{\mathbf{c}}_{\mathrm{s}}\bar{\mathbf{c}}_{\mathrm{s}}^\mathsf{T} \right).
\end{align}
Furthermore, let us define $\mathbf{C}=\left( 0.5\mathsf{diag}\left\{\bar{\mathbf{c}}_{\mathrm{s}}\right\}+\bar{\mathbf{C}}_{\mathrm{i}} \right)^{-1}
\overset{(a)}{=}\mathsf{diag}\big\{\big[\frac{1}{0.5\bar{c}_{\mathrm{s}}[1]+\bar{c}_{\mathrm{i}}^{\mathrm{ext}}},\dots,\frac{1}{0.5\bar{c}_{\mathrm{s}}[M]+\bar{c}_{\mathrm{i}}^{\mathrm{ext}}}\big]\big\}$ where equality $(a)$ holds if the channel is ISI-free. For an eigenvalue $\kappa$ and the corresponding eigenvector $\boldsymbol{\upsilon}$ of a matrix $\mathbf{D}$, equality $\mathbf{D}\boldsymbol{\upsilon}=\kappa \boldsymbol{\upsilon}$ holds. Exploiting this relation, 
$\kappa=\bar{\mathbf{c}}_{\mathrm{s}}^\mathsf{T} \mathbf{C} \bar{\mathbf{c}}_{\mathrm{s}}$ and
$\boldsymbol{\upsilon}= \mathbf{C} \bar{\mathbf{c}}_{\mathrm{s}}$
are an eigenvalue  and the corresponding eigenvector of matrix $\mathbf{D} = \mathbf{C} \bar{\mathbf{c}}_{\mathrm{s}}\bar{\mathbf{c}}_{\mathrm{s}}^\mathsf{T}$ since
\begin{IEEEeqnarray}{ll} 
\kappa \boldsymbol{\upsilon} =  \boldsymbol{\upsilon} \kappa  
= & \mathbf{C} \bar{\mathbf{c}}_{\mathrm{s}} \bar{\mathbf{c}}_{\mathrm{s}}^\mathsf{T}  \mathbf{C} \bar{\mathbf{c}}_{\mathrm{s}} = \mathbf{D} \boldsymbol{\upsilon}
\end{IEEEeqnarray}
holds. Moreover, since $\mathbf{D}$ is a rank-one matrix, it has only \textit{one non-zero} eigenvalue.  Therefore, the aforementioned eigenvalue, $\kappa$, and eigenvector, $\boldsymbol{\upsilon}$, are indeed  $\kappa_{\max}(\mathbf{D})$ and $\boldsymbol{\upsilon}_{\max}(\mathbf{D})$, respectively. In fact, $0.5\kappa_{\max}(\mathbf{D})$ and $\boldsymbol{\upsilon}_{\max}(\mathbf{D})$ are the close-form expressions for $\mathsf{SINR}^{\mathrm{opt}}$ and $\mathbf{f}^{\mathrm{opt}}$ provided in Theorem~\ref{Theo:MF}, respectively. This concludes the proof.

\vspace{0.1cm}

\bibliographystyle{IEEEtran}
\bibliography{Ref_04_05_2017}

\end{document}